\documentclass[%
 reprint,
 amsmath,amssymb,
 prl,
]{revtex4-2}

\usepackage{graphicx}
\usepackage{dcolumn}
\usepackage{bm}

\usepackage{algpseudocode}
\usepackage{algorithm}
\usepackage{physics}
\usepackage{color}
\begin{document}

\preprint{APS/123-QED}

\title{Solving the time-complexity problem and tuning the performance of quantum reservoir computing by artificial memory restriction}

\author{Saud \v{C}indrak}
\email{saud.cindrak@fokus.fraunhofer.de}
\affiliation{Fraunhofer Institute for Open Communication Systems, Berlin, Germany}

\author{Brecht Donvil}%
\affiliation{Institute for Complex Quantum Systems and IQST, Ulm University - Albert-Einstein-Allee 11, D-89069 Ulm, Germany}
\author{Kathy Lüdge}
 \affiliation{%
 Technische Universität
Ilmenau, Institute of Physics, Ilmenau, Germany
}%
\author{Lina Jaurigue}
\affiliation{%
 Technische Universität
Ilmenau, Institute of Physics, Ilmenau, Germany
}%


\date{\today}

\begin{abstract}

Quantum reservoir computing is a computing approach which aims at utilising the complexity and high-dimensionality of small quantum systems, together with the fast trainability of reservoir computing, in order to solve complex tasks.
The suitability of quantum reservoir computing for solving temporal tasks is hindered by the collapse of the quantum system when measurements are made. This leads to the erasure of the memory of the reservoir. Hence, for every output, the entire input signal is needed to reinitialise the reservoir, leading to quadratic time complexity. Overcoming this issue is critical to the hardware implementation of quantum reservoir computing.
We propose artificially restricting the memory of the quantum reservoir by only using a small number inputs to reinitialise the reservoir after measurements are performed, leading to linear time complexity. This not only substantially reduces the number of quantum operations needed to perform timeseries prediction tasks, it also provides a means of tuning the nonlinearity of the response of the reservoir, which can lead to significant performance improvement.
We numerically study the linear and quadratic algorithms for a fully connected transverse Ising model and a quantum processor model. We find that our proposed linear algorithm not only significantly reduces the computational cost but also provides an experimental accessible means to optimise the task specific reservoir computing performance.  

\end{abstract}

\maketitle


\section{Introduction}

The field of quantum computation promises a significant computational speedup over classical computation for certain sets of problems \cite{Nielsen2009}. Machine learning is one such fields where it is know that quantum computers can offer an advantage \cite{BiWi2017}.
Several machine learning tasks have been experimentally realised on quantum systems, some examples are \cite{BrMi2013,LiLi2015,LiHo2017,SaAs2021}, but broad applicability of machine learning on quantum devices is still hindered by the limitations of current quantum processing devices. One of these limitations is the inevitable noise these devices experience. For reservoir computing, a sub-field of machine learning, this noise does not pose a hindrance and could even be a resource \cite{ChNu2020,SuGa2022,GOV21}.

Reservoir computing is a machine learning approach wherein only the output layer is trained \cite{JAE01,MAA02,VER07,APP11}. Due to this simple training scheme it is well suited for hardware implementation, meaning that an input signal is fed into a physical system and the dynamics of that physical "reservoir" are utilised to project the data into a high dimensional latent space. The responses of the reservoir are then sent through a readout layer, which is trained in order to approximate the desired function. 



There are two main avenues of research into quantum reservoirs, either quantum systems whose dynamics are generated by a Hamiltonian $H$ or quantum circuits consisting of several qubits on which unitary operations can be performed. For the former, several studies have been dedicated to the Ising model \cite{FuNa2017,NaFu2019,KUT20,MaPe2021,XIA22}, showing its viability as a reservoir for several benchmark tasks. The authors of \cite{ChNu2020,SuGa2022,PFE22} devised schemes for reservoir computing on a quantum circuit and implemented them on IBM quantum processors. 
A promising avenue of use for quantum reservoir computing is to aid in the measurement of quantum states \cite{AnKh2021,KhaHu2021}.

The output of both types of quantum reservoirs are typically time series of one or two qubit observables. For each output, quantum measurements have to be performed, which poses a significant problem for the physical implementation of quantum reservoir computing \cite{MUJ21a}. With each measurement, the quantum system state collapses and all information about the input signal is lost. Therefore, for each time step of the output, the entire signal up to that point is needed to reinitialise the reservoir. This procedure leads to a time complexity quadratic in the length of the input signal. The authors of \cite{FuNa2017} propose as a solution to perform reservoir computing with nuclear-magnetic-resonance spin ensemble systems \cite{Cory2000,Jo2011}. These large ensembles have the advantage that all copies of the ensemble can be simultaneously controlled such that they all follow the same dynamics. In this way expectation values can be measured with barely any backaction. 
The authors of \cite{MUJ23} investigate the influence of weak measurements and additionally make the observation that due to the fading memory of the reservoir it is not necessary to reset the reservoir using the entire sequence of previous inputs. This can be understood as follows. Typically, information is encoded in the system state of some elements and then letting the closed system evolve in time. Due to the successive over-writing of elements of the quantum reservoir, memory of past inputs is gradually lost. This means that the response of the reservoir is independent of inputs from the distant past and only a finite number of past inputs are needed to reinitialise the reservoir after each measurement.


In this paper we study the influence of artificially restricting the amount of signal inputs after the reservoir is reset by a system measurement. Not only does it reduce the time complexity of the reservoir computing algorithm \cite{MUJ23} but also find that it is possible to tune the task-specific computing performance.
Our proposed approach simultaneously provides an experimentally viable method of tuning the nonlinearity of the quantum reservoir response which addresses the need for task dependent hyperparameter optimisation. Task dependent hyperparameter optimisation is not an issue specific to QRC, but a general issue for reservoir computing, particularly for hardware implemented reservoir computing where the accessible hyperparameters can be restricted and difficult or cumbersome to tune. 

We demonstrate our approach on two simulated quantum reservoirs: a transverse field Ising Hamiltonian and a quantum circuit. In both cases we analyze their performance on the information processing capacity \cite{DAM12} and the Lorenz chaotic attractor. Our proposed algorithm not only addresses the problem of measurement for time series tasks, but also improves the performance for these tasks.

\section{Quantum Measurement for Time Series Predictions}

The usual procedure to experimentally implement time series tasks in quantum reservoirs  involves reinitialising the reservoir with the entire input history. The inputs to this algorithm include a unitary operation ${U}$, the set of observables to be measured $\{{O}_o\}_o$, the initial state $\ket{\Psi_{I}}$, an input series $\textbf{u}=\{u_i\}_{i=1}^M$ of length $M$ and a scheme which encodes the input into a state $\ket{\Psi_{E}(u_i)}$. 
For each time step $i$, the system is initialised to $\ket{\Psi_{I}}$ and the signal up until $u_i$ is fed into the reservoir.  
Afterwards the measurement of the set of observables ${\{O}_o\}_o$ is performed. 
Since the $i$-th input requires $i$ unitary operations, the complexity of this algorithm for a time series of length $M$ is determined by 
\begin{align}
    T_1(M)=\sum_{i=1}^M i = \frac{M(M+1)}{2} \in O(M^2).
\end{align}
For large or continuous time-series ($M \rightarrow \infty$), this approach therefore becomes unfeasible. Current literature on Quantum Reservoir Computing for time-series tasks analyzes the properties of the reservoir using this scheme with quadratic time complexity \cite{FuNa2017,NaFu2019,MaPe2021,ChNu2020,SuGa2022}.


We propose a new scheme for reservoir computing. The scheme is based on the fading memory property usually assumed for reservoir computers \cite{JAE01,MAA02}, and the memory-nonlinearity trade-off which is known to occur in reservoir computing \cite{VER10,BUT13b,INU17}. Qualitatively the fading memory property states that the reservoir forgets inputs far into the past. It can be characterised by the linear contribution to the information processing capacity IPC$_1$. The information processing capacity is a generalization of the linear memory capacity \cite{DAM12} which quantifies the ability of the reservoir to construct nonlinear transforms of all possible combinations of past inputs into the reservoir (see the supplemental material for details) and can be used to predict the performance on certain tasks \cite{HUE22,HUE22a}. 

The scheme we propose here is to only insert the previous $m=n-1$ inputs for each output step of the reservoir $i$, rather than the last $i-1$ inputs. This results in a total of $n$ unitary operation per input $u_i$ and leads to a total of
\begin{align}
    T_2(M)=n\cdot M \in O(M)
\end{align}
unitary operations. This approach is computationally feasible for continuous or large time-series ($M \rightarrow \infty$), as at any given input $u_i$ only $n+1$ unitary operation are required and thus can be computed in real-time.

In the following section we compare the quadratic (QCQA) and the linear complexity quantum algorithms (LCQA) as a function of the reset length $n$. We study two different systems: an Ising model and a quantum circuit. For both systems, we encode our input series in the first qubit $\ket{\Psi_1}$ and set the initial state of the reservoir $\Psi_I$ and the encoding state $\Psi_E(u_i)$ depending on an input $u_i$ to
\begin{align}\label{eq:init}
    \ket{\Psi_{I}} &= \ket{0000}
\end{align}
\begin{align}\label{eq:ecoding}
    \ket{\Psi_{E}(u_i)} &= \sqrt{\frac{1-u_i}{2}}\ket{0}+\sqrt{\frac{1+u_i}{2}}\ket{1}.
\end{align}
\section{Ising Model}

The dynamics of the fully-connected-transverse field Ising model are described by the Hamiltonian
\begin{align}
    {H} = \sum_{i=1,j>i}^{N_S}J_{ij}X_iX_j+\sum_{i=1}^{N_S}hZ_i.
\end{align}
$X_i,~Y_i$ and $Z_i$ are the Pauli matrices of the $i$-th particle and are given by 
\begin{align}
    X_i, Y_i, Z_i = \Big( \bigotimes_{k=1}^{i-1}I_2 \Big)\otimes \sigma_{x,y,z} \otimes\Big( \bigotimes_{k=i+1}^{N_S}I_2 \Big),
\end{align} 
where $\sigma_{x,y,z}$ are the  1-qubit Pauli matrices and $I_2$ the 1-qubit identity operator.
The $J_{ij}$ are the coupling strengths between two particles in the $x$-direction and are sampled from a uniform distribution on the interval $[0.25, 0.75]$, while $h=0.5$ is the coupling strength to an external magnetic field in the $z$-direction. 

We numerically study the Ising model with four qubits $N_S=4$. The outputs of the reservoir are the expectation values $\{\langle Z_i\rangle\}_{i=1}^4$, where $Z_i$ is the $z$ Pauli matrix for the $i$-th qubit. Additionally we perform time multiplexing: each input signal is fed into the reservoir for an evolution time $T$, during which we perform $N_V=30$ measurements, see supplemental material for more details on time-multiplexing. This leads to a total of $4\times30 = 120$ observables or readout nodes for each input step $i$. The unitary time evolution operation is given by
\begin{align}
    {U} = \exp{-i{H}T},
\end{align}
with an evolution time (clock cycle) of $T=20$. 
\begin{figure}[h]
    \centering
    \includegraphics[scale=1]{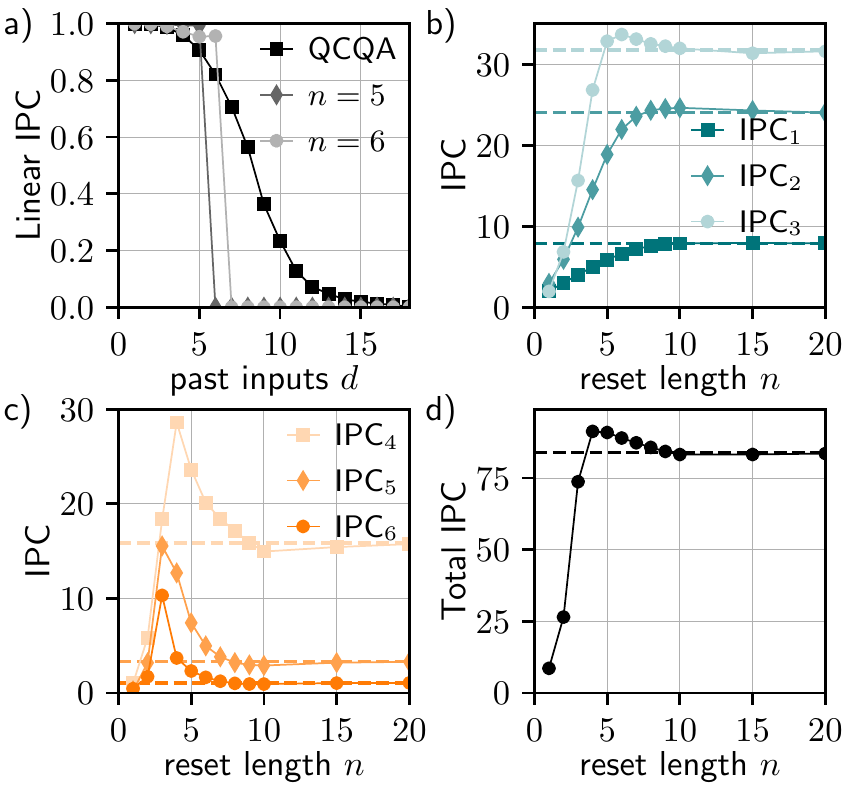}
    \caption{{\bf Ising model:} \textbf{a)} Linear IPCs as a function of the steps into the past $d$ for the LCQA with $n=5,6$ (greys) and for the QCQA (black). Summed IPCs of polynomial orders \textbf{b)} 1 to 3 and \textbf{c)} 4 to 6, and \textbf{d)} the total summed IPC, in dependence of the reset length $n$ using the LCQA. The corresponding QCQA limits are indicated by the dashed lines. We have calculated the standard deviation for ten different realizations of Ising Hamiltonians, where $J_{ij}$ were sampled randomly.}
    \label{fig:IPC_Ising_results}
\end{figure}



For the Ising model, the linear memory capacity as a function of the steps into the past is shown in Fig. 1a. Using the entire history to reset the system, i.e. the QCQA, the memory of this Ising model reservoir fades to zero after approximately 15 steps into the past (black line), meaning that a reset length of $n=15$ should be sufficient to emulate the QCQA. To demonstrate the influence of the reset length in a general and task-independent manner we calculate the 
information processing capacities as a funciton of $n$.  
Figure 1b-c shows the summed IPCs of polynomial order one to six and Fig. 1d shows the total IPC summed over all polynomial orders. In each case the QCQA limit (dashed lines) is reached as $n$ approaches the maximum memory of the reservoir ($n\approx 15$). For very small $n$ the IPCs are decreased due to the artificial memory restriction that is being imposed on the reservoir. However, there is an intermediate range for the reset length $n$ where the total IPC and the IPCs above first order are increased compared with the QCQA limit. This is a new insight, which can be used to substantially reduce the number of quantum operations needed for time series tasks, while simultaneously optimizing the performance. 

To gain more insight into this effect, in Fig. 1a the linear memory (linear IPCs) is plotted for $n=5,6$. 
Here it can be seen that the distribution of the linear IPCs is changed compared with the QCQA case. Although the summed linear IPC (IPC$_1$) is decreased for $n=5,6$, the capacities for the past inputs which can be reconstructed are higher. This increase in the memory is related to the state of the reservoir, which is initially the pure state \eqref{eq:init} and becomes mixed after inputting data. In the supplementary material we show that the data encoded as  
\eqref{eq:ecoding} in a pure state is better remembered than in a mixed state.
This explains the increase in the summed higher order IPCs, since the high order IPCs are composed of inputs from fewer steps into the past, as can be inferred by the small reset lengths $n$ needed to reach the QCQA limit as the polynomial order is increased (see Fig.~\ref{fig:IPC_Ising_results}c).
 
\begin{figure}[h]
    \centering
    \includegraphics[scale=1]{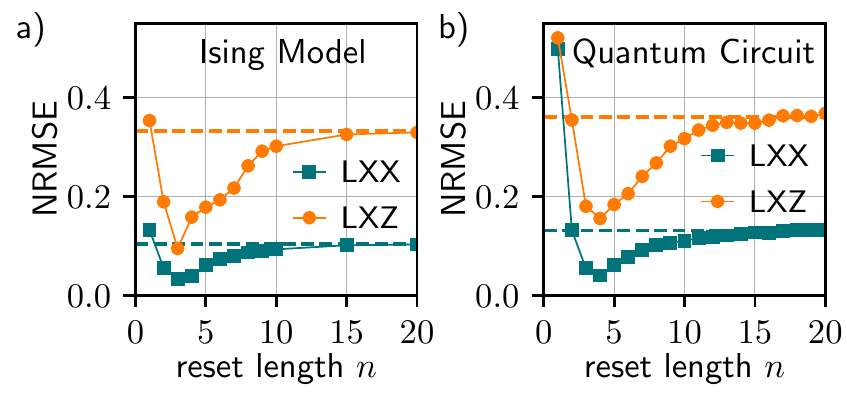}
    \caption{{\bf Lorenz tasks:} NRSME of the LXX (green) and LXZ (orange) tasks for the LCQA as a function of the reset length $n$ using \textbf{a}) the Ising Reservoir and \textbf{b)} the quantum circuit. The QCQA limit is indicated by the dashed lines.}
    \label{fig:IPC_Lorenz_results}
\end{figure}

To demonstrate that the observed increases in the IPCs can translate to improved performance for a time series prediction task, we also calculate two tasks related to the Lorenz chaotic attractor \cite{LOR63}. In both tasks the $x$ variable of the Lorenz system is inserted into the reservoir. The first task (LXX) is to predict the $x$ variable one step ahead. The second task (LXZ) is to cross-predict the $z$ variable one step ahead. (See the supplemental material for details on the tasks.) Figure \ref{fig:IPC_Lorenz_results}a shows the normalised root-mean-squared error (NRMSE) (as defined in the supplemental material) for the LXX and LXZ tasks in dependence of the reset length $n$. For both tasks, compared with the QCQA limit (dashed lines), a lower NRMSE is achieved for small $n$. For this particular reservoir, the minimum NRMSE for both tasks is achieved at $n=3$, which corresponds to the reset length at which IPC$_5$ and IPC$_6$ exhibit a maximum (see Fig.~\ref{fig:IPC_Ising_results}c).


\section{Quantum Circuit}

To demonstrate the universality of our restricted memory approach, the second type of quantum reservoir we consider is an $N$-qubit circuit. Each layer of the circuit consists of two sub-layers of 2-qubit unitary operators $W_j$ and $V_j$ acting on neighbouring qubits.
The unitaries $W_j$ and $V_j$ are of the form
\begin{align}\label{eq:circuit_unitary}
    W_j =&  w_{j,1}\, U_{2j-1,2j}(a_j,b_j,c_j) \,w_{j,2}\nonumber\\ 
    V_j =& v_{j,1}\, U_{2j,2j+1}(d_j,e_j,f_j)\,v_{j,2}
\end{align}
with 
\begin{align*}
    U_{k,l}(a,b,c) = e^{i a\, X_{k}X_{l} +ib\, Y_{k}Y_{l} + icZ_{k}Z_{l}}
\end{align*}
and 
\begin{align*}
    w_{j,k} =& \Big( \bigotimes_{k=1}^{j-2}I_2 \Big)\otimes u_{j,k,1}\otimes  u_{j,k,2}\Big( \bigotimes_{k=j+1}^{N_S}I_2 \Big)\\
    v_{j,k} =& \Big( \bigotimes_{k=1}^{j-1}I_2 \Big)\otimes g_{j,k,1}\otimes  g_{j,k,2}\Big( \bigotimes_{k=j+2}^{N_S}I_2 \Big)
\end{align*}
where the $u_{j,k,l}$ and $g_{j,k,l}$ are single qubit unitaries drawn from the Haar measure \cite{Diestel2014-mh} and $a_i,\, ...\,,\, f_j$ are uniformly drawn from the interval $[-k,h]$. The two sublayers are then defined by 
\begin{align}
    W = \prod_j W_j\quad \textrm{and}\quad V = \prod_j V_j
\end{align}
Recently, the authors of \cite{SuPe2018} showed that if the single qubit unitaries $w,\,v$ are drawn from the Haar distribution, see e.g. \cite{Halmos1974} (Sec. 58), and $k=h$, then by changing $h$ the system undergoes a transition between a localised and an ergodic phase. 
In our case, we draw the parameters randomly from the interval
\begin{equation}
    a_i,\, ...\,,\, f_j\in [0.1,0.2]
\end{equation}

We implement both the QCQA and the LCQA scheme, where the unitary operation ${U}$ is given by repeating the sublayers $W$ and $V$, $N_W=10$ times
\begin{align}
    {U} = (V \, W)^{N_W}
\end{align}
and where we time-multiplex by performing an additional measurement after the application of each $V$ and $W$ layer individually. We use $N=4$ qubits, thus, in total, there are $8\times 10 =80$ outputs of the reservoir for each input.

\begin{figure}[h]
    \centering
    \includegraphics[scale=1]{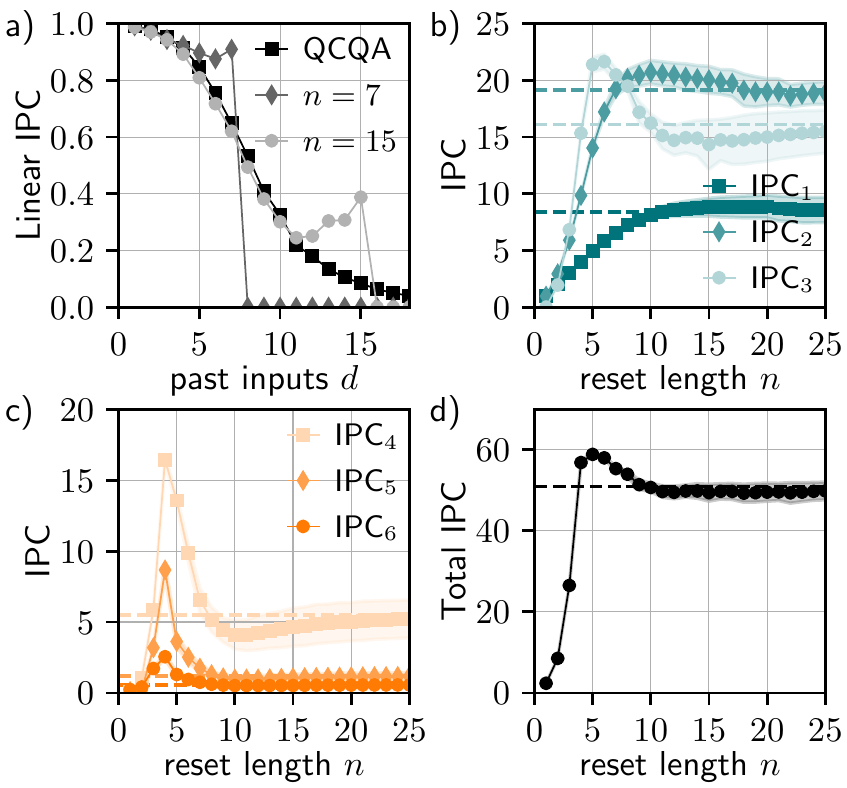}
    \caption{{\bf Quantum Circuit:} \textbf{a)} Linear IPCs as a function of the steps into the past $d$ for the LCQA with $n=7,15$ (greys) and for the QCQA (black). Summed IPCs of polynomial orders \textbf{b)} 1 to 3 and \textbf{c)} 4 to 6, and \textbf{d)} the total summed IPC, in dependence of the reset length $n$ using the LCQA. The corresponding QCQA limits are indicated by the dashed lines. We have calculated the standard deviation for ten different realizations of the input signal and different random parameters $a_j,\hdots f_j$ and single qubit unitaries $w,v$ in \eqref{eq:circuit_unitary}.}
    \label{fig:IPC_Circuit_results}
\end{figure}

Figure \ref{fig:IPC_Circuit_results} shows the various components of the IPC as a function of the reset length $n$. Here we find the same qualitative results as in the Ising model case (see Fig.~\ref{fig:IPC_Ising_results}). The IPCs above first order can be increased with respect to the QCQA limit and for sufficiently large $n$ the QCQA limit is reached. The exact influence  of the reset length on the distribution of the IPCs depends on the dynamics of the reservoir, as can be seen by the differences between Fig. 1 and Fig.~\ref{fig:IPC_Circuit_results}.



For the quantum circuit, optimisation of the reset length also leads to an improvement in the performance of the LXX and LXZ tasks, as shown in Fig.~\ref{fig:IPC_Lorenz_results}b. Here the best performances occur for $n=4$, which corresponds to the maximum IPC$_5$ and IPC$_6$ for the quantum circuit.


\section{Conclusion}

In this work, we presented the Linear Complexity Quantum Algorithm (LCQA), for quantum reservoir computing. The algorithm successfully reduces the time-complexity of quantum reservoir computing for time series tasks from quadratic to linear, thus making physical implementations for long time series feasible. Beyond this, we have demonstrated that by artificially restricting the memory of the reservoir, the nonlinear response can be tuned and a further reduction in the required number of quantum operations can be achieved.

We have compared our new LCQA approach to the established QCQA on a fully connected Ising chain and a quantum processor reservoir computer. We found that LCQA outperforms the currently utilized QCQA scheme both in the information processing capacity and in Lorenz time series prediction tasks.

The proposed approach allows the nonlinearity of the reservoir response to be tuned at the expense of the linear memory. For tasks requiring greater memory, the LCQA can be supplemented with memory augmentation methods on the input or the output of the reservoir, such as those presented in \cite{MAR19a,DEL21a,JAU21a,CAR22a}.

Our findings from the evaluation of the LCQA scheme using the quantum circuit indicate that this algorithm shows great potential for the hardware implementation of quantum reservoir computing. Not only to reduce the time needed to perform computations, but to improve the performance.
The proposed LCQA algorithm presents a promising avenue for further research and development in the field of quantum computing, quantum reservoir computing, and quantum machine learning. 

\begin{acknowledgments}
L. J. acknowledges funding from the Deutsche Forschungsgemeinschaft (DFG), grant number
LU 1729/3-1.
\end{acknowledgments}

\bibliography{lit_2}


\end{document}


\preprint{APS/123-QED}

\title{Solving the time-complexity problem and tuning the performance of quantum reservoir computing by artificial memory restriction}

\author{Saud \v{C}indrak}
\email{saud.cindrak@fokus.fraunhofer.de}
\affiliation{Fraunhofer Institute for Open Communication Systems, Berlin, Germany}

\author{Brecht Donvil}%
\affiliation{Institute for Complex Quantum Systems and IQST, Ulm University - Albert-Einstein-Allee 11, D-89069 Ulm, Germany}
\author{Kathy Lüdge}
 \affiliation{%
 Technische Universität
Ilmenau, Institute of Physics, Ilmenau, Germany
}%
\author{Lina Jaurigue}
\affiliation{%
 Technische Universität
Ilmenau, Institute of Physics, Ilmenau, Germany
}%

\collaboration{CLEO Collaboration}

\date{\today}

\maketitle

\tableofcontents

\newpage

\section{Parameters}
The following table contains the parameters for the simulation. 
\begin{table}[H]
\centering
\begin{tabular}{|l|l|}
\hline
\textbf{Parameter}                                                                              & \textbf{Value}                               \\ \hline
init steps                                                                                      & 10 000                                       \\ \hline
training steps                                                                                  & 50 000                                       \\ \hline
testing steps                                                                                   & 5 000                                        \\ \hline
\begin{tabular}[c]{@{}l@{}}regularization parameter \\ quantum circuit, tikanov\end{tabular}    & $10^{-2}$                                    \\ \hline
\begin{tabular}[c]{@{}l@{}}regularization parameter \\ Ising model, noise\end{tabular}          & $10^{-6}$                                    \\ \hline
\begin{tabular}[c]{@{}l@{}}sampling, quantum circuit\\ haar-distrubution with para\end{tabular} & $a_j,.,f_j \in [0.1, 0.2]$                  \\ \hline
\begin{tabular}[c]{@{}l@{}}sampling, Ising model\\ coupling $J$\end{tabular}                    & $J_{ij}$ uniformly sampled on $[0.25, 0.75]$, $h_i=5$ \\ \hline
time evolution for Ising model                                                                  & $T=20$                                       \\ \hline
number of qubits                                                                                & $N_Q = 4$                                        \\ \hline
\end{tabular}
\caption{Parameters for simulations.}
\label{tab:parameters}
\end{table}

\section{Reservoir Computing}

A reservoir computer (RC) can be understood as a recurrent neural network, where only the output weights $\textbf{W}^{out}$ are trained. The internal nodes are either chosen randomly or can not be influenced directly. The input and output of the network can be weighed with pre- and post-processing, respectively. The two main benefits of a physical RC are its faster training speed and faster computation time. Faster training is based on the fact, that the internal weights are not trained. 
Faster computation time can be explained with the fact that the main processing of the network is done by a physical system and not with neurons on a computer. 

Assume an input series $\textbf{u}=(u(t_1), u(t_2),..)$ and $N_S$ readout nodes. At each time $t_i$ the $i-$th input $u_i$ is introduced into the system and the output $(s_1(t_i),\hdots s_{N_S}(t_i))$ is obtained. The target sequence at each $t_i$ is  ${y}^{\operatorname{targ}}(t_i)$ and the estimate by the reservoir is obtained by taking a linear combination of the output signal weighed by the vector $W^{\text {out }}$
\begin{align}
    y(t_i) = \sum_j W^{\text {out }}_j s_j(t_i).
\end{align}
The distance function $\epsilon = \sum_{i}\left({y}^{\operatorname{targ}}\left(t_{i}\right)-{y}\left(t_{i}\right)\right)^{2}$ will be used. For such a distance function the optimum weights are obtained by
\begin{align}
W_{o p t}^{\text {out }}&=\arg \min _{W^{\text {out }}} \sum_{i}\left({y}^{\operatorname{targ}}\left(t_{i}\right)-{y}\left(t_{i}\right)\right)^{2} ={S}^{-1} \textbf{Y}^{\textbf{targ}}
\end{align}
where ${Y}_{i,j}^{\textbf{targ}}=y_j^{\textbf{targ}}(t_i)$ and ${S}^{-1}$ is the Moore-Penrose inverse of the state matrix ${S}_{i, j}=s_{j}\left(t_{i}\right)$, where $j\in \{1,2,..,J\}$ is the number of readout nodes. \\
\leavevmode \\
\textbf{Time multiplexing}

To increase the output dimension of the state matrix, the $N_S$ output nodes can be multiplexed. Assume a total of $n$ inputs, where the $i$-th input is introduced into the system. Rather than just computing one output $s_j(t_i+T)$, the output for $N_V$ times $t_i+k\theta$ can be computed, where $j\in \{1,2,..,N_S\}$, $k\in \{1,2,..,N_V\}$, $t_i=iT$ such that $N_V\theta = T$ holds. The following state matrix can be constructed.
\begin{align}
S &= \begin{pmatrix}
s_1(t_1) &   ..  & s_{N_s}(t_1) & z_1(t_1+\theta)   & ..  & s_{N_s}(t_1+\theta)&.. & s_1(t_1 + T)   & ..  & s_{N_s}(t_1+ T)\\ 
s_1(t_2) &  ..  & s_{N_s}(t_2) & s_1(t_2+\theta)  & ..  & s_{N_s}(t_2+\theta)&.. & s_1(t_0 + T) & ..  & s_{N_s}(t_0+ T))\\
 :& : & :  & : &:&:&:&:&:&:\\ 
s_1(t_n)  & ..  & s_{N_s}(t_n)) & s_1(t_n+\theta) &  ..  & s_{N_s}(t_n+\theta))&.. & s_1(t_n + T) &  ..  & s_{N_s}(t_n+ T))\\ 
\end{pmatrix} 
\end{align}\\

\begin{figure}[h]
	\hspace*{-0.5 cm}
	\centering
	\includegraphics[scale=0.25]{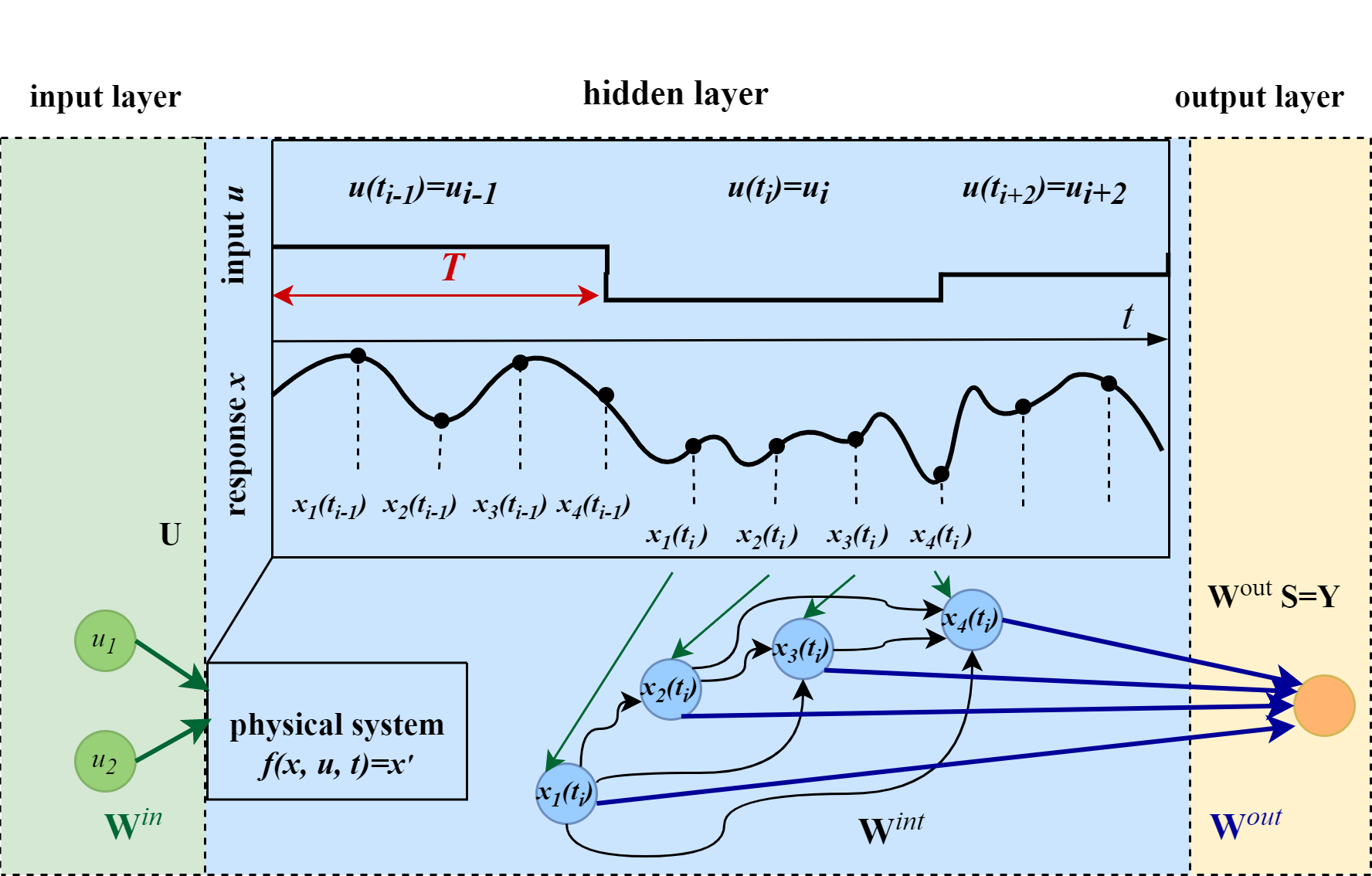}
	\caption[Sketch of reservoir computer]{A sketch of a reservoir computer. Here a two-dimensional input at time $t$  $\textbf{U}=(u_1, u_2)$ is given. The input nodes $u_i$ are then connected by the input weights $\textbf{W}^{\text{in}}$ to a physical system. 
	The evolution over time of state $x$ is governed by a function, that depends on the input $u$, the time $t$ and the state of the system $x$. An sketch of a possible evolution is shown in the upper part of the hidden layer segment. The $x$-axis denotes the time and the upper plot is denotes the input $u$. Three different inputs $u(t_{i-1})$, $u(t_{i})$ and $u(t_{i+11})$ are shown in this plot. The difference in time between two inputs is referred to as clock cycle $T=t_{i+1}-t_{i}$. The lower graph shows a response $x$ of the physical system in regards to the input $u$. To obtain the virtual nodes $x_j(t_i)$ the system is sampled $N_V$ times for each input. This sketch shows four virtual nodes, which are visualized by the dashed lines. These measurements can be understood as virtual nodes $x_j(t_i)$ that are coupled to each other by the internal weights $\textbf{W}^{\text{int}}$. The coupling of the internal weights depends on the dynamics of the system, which in turn depend of the input, the time and the state. A subset of the virtual nodes is used as readout nodes $s_i$ and the state matrix $\textbf{S}$ can be constructed.
	This example shows a $\textbf{S}(t_i) = [s_1(t_i), s_2(t_i), s_3(t_i), s_3(t_i)]= [x_1(t_i), x_2(t_i), x_3(t_i), x_4(t_i)]$. By multiplying the \textbf{S} by the output weights $\textbf{W}^{\text{out}}$ the output $\textbf{Y}$ can be obtained.}
\label{fig:1_3}
\end{figure}

Figure \ref{fig:1_3} shows a sketch of a reservoir computing scheme. A input layer with two inputs is fed into a physical system with the input weights $\textbf{W}^{in}$. Often systems of the form $f(x,u,t)=dx/dt$ are used, where $x$ denotes the response of the system, $u$ is the input vector and $t$ is the time. In dependence of the input $u$ the system response $x$ evolves over time. The time between two inputs is called the clock cycle $T$ and is shown in red in the sketch. Virtual nodes are generated by sampling the response for multiple times in the interval $t\in [iT, (i+1)T]$. The number of measurements is called the number of virtual nodes $N_V$ and the time between two virtual nodes is called virtual node separation $\theta = T/N_V$. With this, we can sketch the virtual nodes as nodes $\textbf{X}(t_i) = (x_1(t_i),.., x_4(t_i))$ and plot the physical interaction of the system, see Fig. \ref{fig:1_3}. A subset of the virtual nodes is taken as readout nodes. In this sketch all virtual nodes are taken as readout nodes $\textbf{S}(t_i)=\textbf{X}(t_i)$. The internal weights $\textbf{W}^{int}$ of a physical reservoir can not be directly changed and can only be influenced by changing the physical system itself. Lastly the constructed state matrix will be multiplied with the output weights $\textbf{W}^{out}$ to obtain the output $\textbf{Y}$.\\

\section{Tasks}

\subsection{Information Processing Capacity}
The information processing capcaity IPC is often used as a measure to quantify the computational power of a RC. We will also try to give some examples to make the combinational problems clear.\\
Additionally the Legendre polynomials will be introduced as an orthogonal basis, such that the information processing capacity IPC can be defined in a build-up way. \\\\

\leavevmode\\
\textbf{Definition 1: Legendre polynomials}\\
The Legendre polynomials $l_n:[-1,1]\rightarrow [-1,1]$ are defined by:
\begin{align}
&l_0(x) = 1 \nonumber\\
&l_1(x) = x \nonumber\\
&l_2(x) = \frac{1}{2}(3x^2-1) \nonumber\\
&l_{n+1}(x) = \frac{2n+1}{n+1}x l_n(x)-\frac{n}{n+1}l_{n-1}(x),~\text{for } n\geq 2
\end{align}
They are orthogonal under the scalar product: 
\begin{align}
&\int_{-1}^1 l_i(x)l_j(x)dx=\delta_{i,j}
\end{align}
\leavevmode\\
\textbf{Definition 2: Capacity}\\
Assume a RC with a state matrix $\textbf{S}$ and the output $\textbf{y}=\textbf{S}\textbf{W}^{out}$ with the target $\textbf{y}^{\text{targ}}$. The capacity  to calculate the target $\textbf{y}^{\text{targ}}$ is defined as
\begin{align}
&C(\textbf{y}, \textbf{y}^{\text{targ}}) = \frac{\text{cov}(\textbf{y}, \textbf{y}^{\text{targ}})}{\sigma^2(\textbf{y})\sigma^2(\textbf{y}^{\text{targ}})}.
\end{align}
$\text{cov}(\textbf{y}, \textbf{y}^{\text{targ}})$ is the covariance between the output $\textbf{y}$ and target $\textbf{y}^{\text{targ}}$, $ \sigma(\textbf{y})$ and $\sigma(\textbf{y}^{\text{targ}})$ are the standard deviations of the output and target.\\
\leavevmode\\
\textbf{Definition 3: Information processing capacity}\\
The information processing capacity tries to classify the computational power of a RC by examining how well the system maps a set of orthogonal functions. Here the Legendre polynomials $l_i$ are used as a set of orthogonal functions, see Def. 1.2. Two combinational problems arise. First all possible combinations of the Legendre polynomials with the same polynomial order have to be calculated, as described in Def. 3.1. The second problem is to calculate all admitted combinations of the past inputs, see Def. 3.3.\\

\textbf{Definition 3.1 : Combinations of the Legendre polynomials}\\
Let $k$ be the order of a polynomial and let $\mathcal{D}^k=[d_1,d_2,..,d_q]$ denote a tuple with $0<d_j\leq k $ and $d_a\leq d_b$ for $a<b$, such that $\sum_{j=1}^q d_j = k$ holds true. For each order $k$ there exists a finite number of tuples $N_k$ with these conditions. The $o$-th tuple is addressed by $\mathcal{D}^k_o=[d_{o,1}, d_{o,2}..,d_{o,q}]$, where $o\in \{1,2,..,N_k\}$ holds. 
Let $M^k_o$ be the length of $\mathcal{D}^k_o$. For example the $L$-th tuple of the $K$-th order is defined as $\mathcal{D}^K_L=[d_{L,1}, d_{L,2},d_{L,3}]$, with $M^K_L=3$ and $d_{L,1}+d_{L,2}+d_{L,3}=K$.
The $o$-th polynomial of $p$-th order is defined as
\begin{align}
    &p^k_o = \prod_{j=1}^{M^k_0} l_{d_{o,j}},
\end{align} 
where $l_{d_{o,j}}$ is the $d_{o,j}$-th Legendre polynomial.
We will give an example for the polynomial of order $k=3$. There exist three combinations such that the above conditions are met. Which are $\mathcal{D}^3_1=[d_{1,1}]=[3]$, $\mathcal{D}^3_2=[d_{2,1}, d_{2,2}]=[1,2]$ and $\mathcal{D}^3_3=[d_{3,1}, d_{3,2}, d_{3,3}]=[1,1,1]$. From these tuples the lengths $M^3_1=1$, $M^3_2=2$ and $M^3_3=3$ can be calculated. The polynomials of the third order are given by
\begin{align}
    p^3_1 &= \prod_{j=1}^{M^3_1} l_{d_{1,j}} = l_3 \nonumber\\
    p^3_2 &= \prod_{j=1}^{M^3_2} l_{d_{21,j}} = l_1l_2 \nonumber\\
    p^3_1 &= \prod_{j=1}^{M^3_3} l_{d_{3,j}} = l_1 l_1 l_1.
\end{align}
Assume the input vector $\textbf{u}_{all} = (u_{-b+1}, u_{-b+2}, .., u_{0}, u_{1},u_{2} .., u_{n})$. This vector can be split into the vector for the state matrix $\textbf{u} = (u_{1},u_{2} .., u_{n})$ and the vector prior  \\
$\textbf{u}_{prior}=(u_{-b+1}, u_{-b+2}, .., u_{0})$. $u(m)$ and $u(m-i_1)$ denote the $m$-th input and the shifted input by $i_1$ time steps into the past, where $-b\leqm\leq n$ and $i_1\leq b$ holds. The shift of the system vector $\textbf{u}$ by $i_1$ time steps into the past will be addressed with $\textbf{u}(i_1) = (u_{1-i_1},u_{2-i_1} .., u_{n-i_1})$. The $a$-th Legendre polynomial of the input vector $\textbf{u}$ will be denoted as \\
$\textbf{l}_a(\textbf{u})=(l_a(u_1), l_a(u_2),..,l_a(u_n))$ and the shifted $a$-th Legendre polynomial will be defined as 
\begin{align}
    \textbf{l}_a(i_1) = \textbf{l}_a(\textbf{u}(i_1)) = (l_a(u_{1-i_1}), l_a(u_{2-i_1}),..,l_a(u_{n-i_1}))
\end{align}
The point-wise multiplication between two vectors $\textbf{a}=(a_1, a_2, ..,a_n)$ and $\textbf{b}=(b_1, b_2, ..,b_n)$ results in a new vector $\textbf{c}=\textbf{a}\odot \textbf{b} = (a_1, a_2, ..,a_n)\odot (b_1, .., b_n) = (a_1b_1, a_2b_2,..,a_nb_n)$. 

The polynomials for different steps in time are calculated by
\begin{align}
    &\textbf{p}^k_o(i_1,..,i_q) = \bigodot_{j=1}^{M^k_o} \textbf{l}_{d_{o,j}}(i_j).
\end{align} 
As an example the polynomials of third order with regards to the past inputs $i_j$ are given by:
\begin{align}
    \textbf{p}^3_1(i_1) &= \bigodot_{j=1}^{M_1^3} \textbf{l}_{d_{1,j}} (i_j) = \textbf{l}_3(i_1) \nonumber\\
    \textbf{p}^3_2(i_1,i_2) &= \bigodot_{j=1}^{M_2^3} \textbf{l}_{d_{1,j}}(i_j) = \textbf{l}_1(i_1)\textbf{l}_2(i_2) \nonumber\\
    \textbf{p}^3_3(i_1, i_2, i_3) &= \bigodot_{j=1}^{M_3^3} \textbf{l}_{d_{1,j}}(i_j) = \textbf{l}_1(i_1) \textbf{l}_1(i_2) \textbf{l}_1(i_3).
\end{align}
\leavevmode\\

\textbf{Definition 3.2: Legendre memory accuracy}\\
Assume that the target function is of the form $\textbf{y}^{\text{targ}}(i_1,i_2,..) = \textbf{p}^k_l(i_1,i_2,..)$ for one tuple of $(i_1,i_2,..)$. The capacity of such a target is defined as 
\begin{align}
C_{k,l}^{\text{comb}}(i_1,i_2,..)&=\frac{\text{cov}(\textbf{y}, \textbf{y}^{\text{targ}}(i_1,i_2,..))}{\sigma^2(\textbf{y})\sigma^2(\textbf{y}^{\text{targ}}(i_1,i_2,..)} \nonumber \\
&=\frac{\text{cov}(\textbf{y}, \textbf{p}^k_l(i_1,i_2,..))}{\sigma^2(\textbf{y})\sigma^2(\textbf{p}^k_l(i_1,i_2,..))}.
\end{align} 

\textbf{Definition 3.3: Combinations of past inputs}\\
Valid combinations of $(i_1,i_2,..)$ are those that comply with the following two conditions:
\begin{enumerate}
    \item Only combinations are allowed where all $i_1,~i_2,..$ are unequal to each other ($i_a\neq i_b$, if $a\neq b$).
    \item The second condition is best understood with an example. \\
    Let $\textbf{p}^7_1(i_1, .., i_4)=\textbf{l}_1(i_1)\textbf{l}_1(i_2)\textbf{l}_1(i_3)\textbf{l}_2(i_4)\textbf{l}_2(i_5)$. There are three Legendre polynomials of first order with indices $(i_1, i_2, i_3)$ and two Legendre polynomial of second order with indices $(i_4, i_5)$. The second condition is on the indices where the Legendre polynomials are of the same order. In these subsets only decreasing indices are allowed. Meaning $i_1 > i_2 > i_3$ and $i_4 > i_5$. It is noted that $i_4$ can be greater than $i_3$. One valid combination is $(i_1,i_2,i_3, i_4, i_5)= (5,4,2,3,1)$.
    
\end{enumerate}

\textbf{Example 3.4:}\\
Let $(i_1, i_2, .., i_n)\in \text{comb}$, where the set \text{comb} consists of all valid combinations. The combinations for $1\leq i_j\leq 4$ for the first three orders of polynoms and one realization of fourth order are given by:

\begin{enumerate}
    \item {First order polynomials:
    \begin{align}
        \cdot~ &p_{1,1}(i_1)=l_1(i_1) \nonumber \\
        &~~~ i_1 \in \text{comb}_{1,1}=\{1,2,3,4\}
    \end{align}}
    
    \item Second order polynomials:
    \begin{align}
        \cdot~ &p_{2,1}(i_1)=l_2(i_1) \nonumber \\
        &~~~i_1 \in \text{comb}_{2,1}=\{1,2,3,4\}  \nonumber\\
        \cdot ~&p_{2,2}(i_1, i_2)=l_1(i_1)l_1(i_2) \nonumber \\
        &~~~(i_1, i_2) \in \text{comb}_{2,2}=\{(2,1), (3,1), (4,1), (3,2), (4,2), (4,3)\}
    \end{align}
    
    \item Third order polynomials:
    \begin{align}
        \cdot ~&p_{3,1}(i_1)=l_3 (i_1) \nonumber \\
        &~~~i_1 \in \text{comb}_{3,1}=\{1,2,3,4\} \nonumber \\
        \cdot~ &p_{3,2}(i_1, i_2)=l_1(i_1)l_2(i_2) \nonumber \\
        &~~~(i_1, i_2) \in\text{comb}_{3,2} =  \{(2, 1), (3, 1), (4, 1), (1, 2), (3, 2), (4, 2), \nonumber \\
        &~~~~~~~~~~~~~~~~~~~~~~~~~~~~~~(1, 3), (2, 3), (4, 3), (1, 4), (2, 4), (3, 4)\} \nonumber \\
        \cdot~ &p_{3,3}(i_1, i_2, i_3)=l_1(i_1)l_1(i_2)l_1(i_3) \nonumber \\
        &~~~(i_1, i_2, i_3) \in \text{comb}_{3,3} = \{(3,2,1), (4,2,1), (4,3,1), (4,3, 2) 
    \end{align}
    
    \item one fourth order polynomial:
    \begin{align}
        \cdot~ &p_{4,4}(i_1,i_2,i_3)=l_1(i_1)l_1(i_1)l_2(i_2) \nonumber \\
        &~~~(i_1, i_2, i_3)\in \text{comb}_{4,4}=\{(3, 2, 1), (4, 2, 1), (4, 3, 1), (3, 1, 2), (4, 1, 2), (4, 3, 2), \nonumber\\
        &~~~~~~~~~~~~~~~~~~~~~~~~~~~~~~~~~~(2, 1, 3), (4, 1, 3), (4, 2, 3), (2, 1, 4), (3, 1, 4), (3, 2, 4)
    \end{align}
\end{enumerate}
This example should help the reader to better understand how the $i_j$ are picked for each polynom. In the next step a sum over all valid combinations $\text{comb}_{k,l}$ of $C_{k,l}^{\text{comb}}$ is taken, to get the capacity $C_{k,l}$ in regards to the target polynomial $\textbf{p}_l^k$. 
\begin{align}
    &C_{k,l} = \sum_{\text{comb}_{k,o}} C_{k,l}^{\text{comb}}(i_1,i_2,..)
\end{align}\\

\textbf{Definition 3.5: Information processing capacity (\text{IPC})}\\
As described in Def. 3.1 $N_k$ different polynomial combinations of degree $k$ exist, where $o\in \{1,2,..,N_k\}$ is the index for these polynomials. The $k$-th order capacity $C_k$  is given as the sum over all polynomial combinations of degree $k$. 
\begin{align}
&\text{IPC}_{k} = \sum_{l} C_{k,l}
\label{eq:IPC_k}
\end{align}
The  \textbf{information processing capacity (IPC)} is obtained from the sum over all IPC$_k$.
\begin{align}
&\text{IPC}=\sum_{k}\text{IPC}_k
\label{eq:IPC}
\end{align}

The information processing capacity IPC is often used as a measure to describe how non-linear behaviour of the past inputs can be described. It should be noted that further restraints on the combinations need to be made, as the system exhibits the fading memory property. This means that $lim_{i_1\rightarrow \infty}\rightarrow 0$ and a termination conditions needs to be added. 
Additionally IPC is upper bounded $C\leq N_R$, where $N_R$ is the number of readout nodes \cite{DAM12}.

\subsection{Lorenz Task}
Another time series prediction task is the Lorenz Task, which is based by the evolution of a chaotic Lorenz attractor\cite{LOR63}. The Lorenz attractor dynamics are governd by \ref{eq:Lor}.
\begin{align}
    \dot{X} &= a(Y-X) \nonumber \\
    \dot{Y} &= X(b-Z)-Y \nonumber \\
    \dot{Z} &= XY - cZ
    \label{eq:Lor}
\end{align}
For time series prediction, the system variables will be discretized, such that the series $X_n = X(n\Delta t),~Y_n = Y(n\Delta t)$ and $Z_n = Z(n\Delta t) $ are constructed. In this thesis the Lorenz $XX$ and Lorenz $XZ$ tasks are used as target functions. The Lorenz $XX$ task tries to predict the future of the $X_n$ variable, where the reservoir is driven with the $X_n$. The Lorenz $XZ$ task gets $X_n$ as an input and tries to predict $Z_n$. The most common used parameters in research are  $a=10, ~b=28,~c=8/3$ with the discretization $\Delta t=0.1$, which will be used here.\\

The normalized root mean squared error (NRMSE)  between  $\textbf{y}=(y_1,y_2,..,y_N)$ and $\textbf{y}^{\text{targ}}=(y_1^{targ},y_2^{targ},..,y_N^{targ})$ is defined as
\begin{align}
    \text{NRMSE}= \sqrt{\frac{\sum_{i=1}^N (y_i-y_i^{\text{targ}})^2}{N\text{var}(\textbf{y}^{\text{targ}})}} = \sqrt{1-
C(\textbf{y}, \textbf{y}^{\text{targ}})}
\end{align}

\section{Quantum Reservoir Computing}

\subsection{Quadratic Scheme}

\begin{algorithm}[h]
\caption{Quadratic Complexity Quantum Algorithm  (\textbf{QCQA}).}\label{alg:alg1}
\begin{algorithmic}[1]
\Require ${U},~\ket{\Psi_{I}}, \ket{\Psi_{E}(u_m)}, \{{O}_o\}_o$
\Require $\textbf{u} \gets [u_1,u_2,..,u_M]$ \Comment{Input series}
\State $i \gets 1$
\While{$i \leq M$}
\State $\rho(0) \gets  \ket{\Psi_{I}}   \bra{\Psi_{I}}$
\State $j \gets 1$
\While{$j\leq i$}

\State $\tilde{\rho}(j) = \ket{\Psi_{E}(u_j)}  \bra{\Psi_{E}(u_j)}\otimes \textrm{Tr}_1(\rho(j-1))$
\State $\rho(j) = {U} \tilde{\rho}(j-1){U}^\dagger$
\State $j \gets j+1$
\EndWhile
\State $S_{m,o}\gets \text{Tr}[\rho(t_m){O}_o]$  \Comment{collapse of state}
\State $i \gets i+1$
\EndWhile
\State \Return $S$
\label{alg:Evol2}

\end{algorithmic}
\end{algorithm}
In quantum reservoir computing the collapse of the state when measuring is usually not considered. Algorithm 1 shows how time series tasks in quantum reservoir comuting would be implemented on a physical quantum reservoir, which we will call the Quadratic Complexity Quantum Algorithm or QCQA. The inputs to this algorithm include a unitary operation ${U}$, the set of observables to be measured $\{{O}_o\}_o$, the initial state $\ket{\Psi_{I}}$, an input series $\textbf{u}=\{u_i\}_{i=1}^M$ of length $M$ and a scheme which encodes the input into a state $\ket{\Psi_{E}(u_i)}$. The first while loop, indexed by $i$ iterates over the input series.
At first $u_1$ is inserted into the system, evolved and then measured, such that the state collapses afterwards. To insert $u_2$ into the system, the state prior the measurement of after the first input has to be reconstructed. This is achieved by inserting $u_1$ into the system, evolving the system and only then inserting $u_2$ into the system, evolving it and then measuring it. Therefore for the tenth input $u_{10}$ all prior inputs $u_1,..,u_9$ have to be inserted. Only then is the measurement on a set of observables ${\{O}_o\}_o$ performed. 

Since the $i$-th input requires $i$ unitary operations, the complexity of this algorithm for a time series of length $M$ is determined by 
\begin{align}
    T_1(M)=\sum_{i=1}^M i = \frac{M(M+1)}{2} \in O(M^2).
\end{align}
For large or continuous time-series ($M \rightarrow \infty$) QCQA becomes unfeasible. Current literature on Quantum Reservoir Computing for time-series tasks analyzes the properties of the reservoir using this scheme with quadratic time complexity \cite{FuNa2017,NaFu2019,MaPe2021,ChNu2020,SuGa2022}.

\subsection{Linear Scheme}
We will now discuss a linear time complexity method that can additionally increase performance. The method is based on the fading memory property, which is usually assumed for reservoir computers. Qualitatively, the fading memory property states that the reservoir forgets inputs far into the past. It can be characterized by the linear contribution to the information processing capacity IPC$_1$. 

This method relies on only inserting the prior $n-1$ inputs for each output step of the reservoir $i$, rather than the last $i-1$ inputs. We show the adjusted algorithm in Algorithm 2, which will be called LCQA. This results in a total of $n$ unitary operation per input $u_i$ and leads to a total of
\begin{align}
    T_2(M)=n\cdot M \in O(M)
\end{align}
unitary operations. This approach is computationally feasible for continuous or large time-series ($M \rightarrow \infty$), as at any given input $u_i$ only $n$ unitary operation are required and thus can be computed in real-time.

\begin{algorithm}[h]
\caption{Linear Complexity Quantum Algorithm (\textbf{LCQA}).}\label{alg:alg2}
\begin{algorithmic}
\State $\rho(0) \gets  \ket{\Psi_{I}}   \bra{\Psi_{I}}$ 
\State $k \gets 0$
\While{$k\leq n$}
\State $j\gets i-n+k$
\State $k \gets k+1$
\State $\tilde{\rho}(j) = \ket{\Psi_{E}(u_j)}  \bra{\Psi_{E}(u_j)}\otimes \textrm{Tr}_1(\rho(j-1))$
\State $\rho(j) = {U} \tilde{\rho}(i){U}^\dagger$
\EndWhile
\label{alg:Evol2}
\end{algorithmic}
\end{algorithm}

In the following section we compare  QCQA and LCQA algorithms in dependence of the reset length $n$. We study two different systems: first an Ising model and then a quantum circuit both consisting of $N_S = 4$ qubits. For both systems, we encode our input series in the first qubit $\ket{\Psi_1}$ and set the initial state of the reservoir $\Psi_I$ and the encoding state $\Psi_E(u_i)$ depending on an input $u_i$ to
\begin{align}
    \ket{\Psi_{E}(u_i)} &= \sqrt{\frac{1-u_i}{2}}\ket{0}+\sqrt{\frac{1+u_i}{2}}\ket{1}.
\end{align}
In the next chapter we will discuss different starting states $\ket{\Psi_I}$.

\section{Simulations with different initial Starting State}

\begin{figure}[h]
	\hspace*{-0.5 cm}
	\centering
	\includegraphics[scale=0.9]{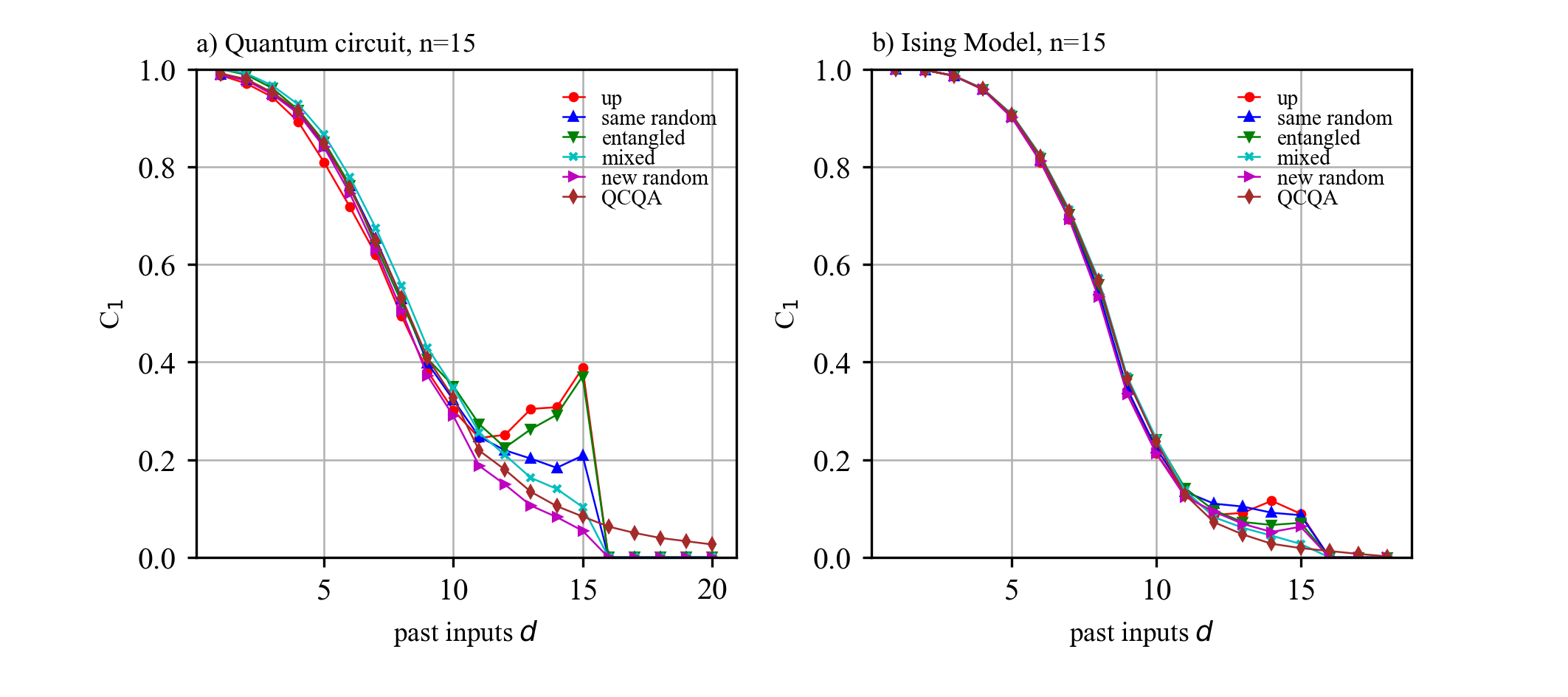}
	\caption[Img 1]{Capacity for the \textbf{a)} quantum circuit \textbf{b)} and Ising model with an reset length of $n=15$. }
\label{fig:S2}
\end{figure}

\begin{figure}[h]
	\hspace*{-0.5 cm}
	\centering
	\includegraphics[scale=0.9]{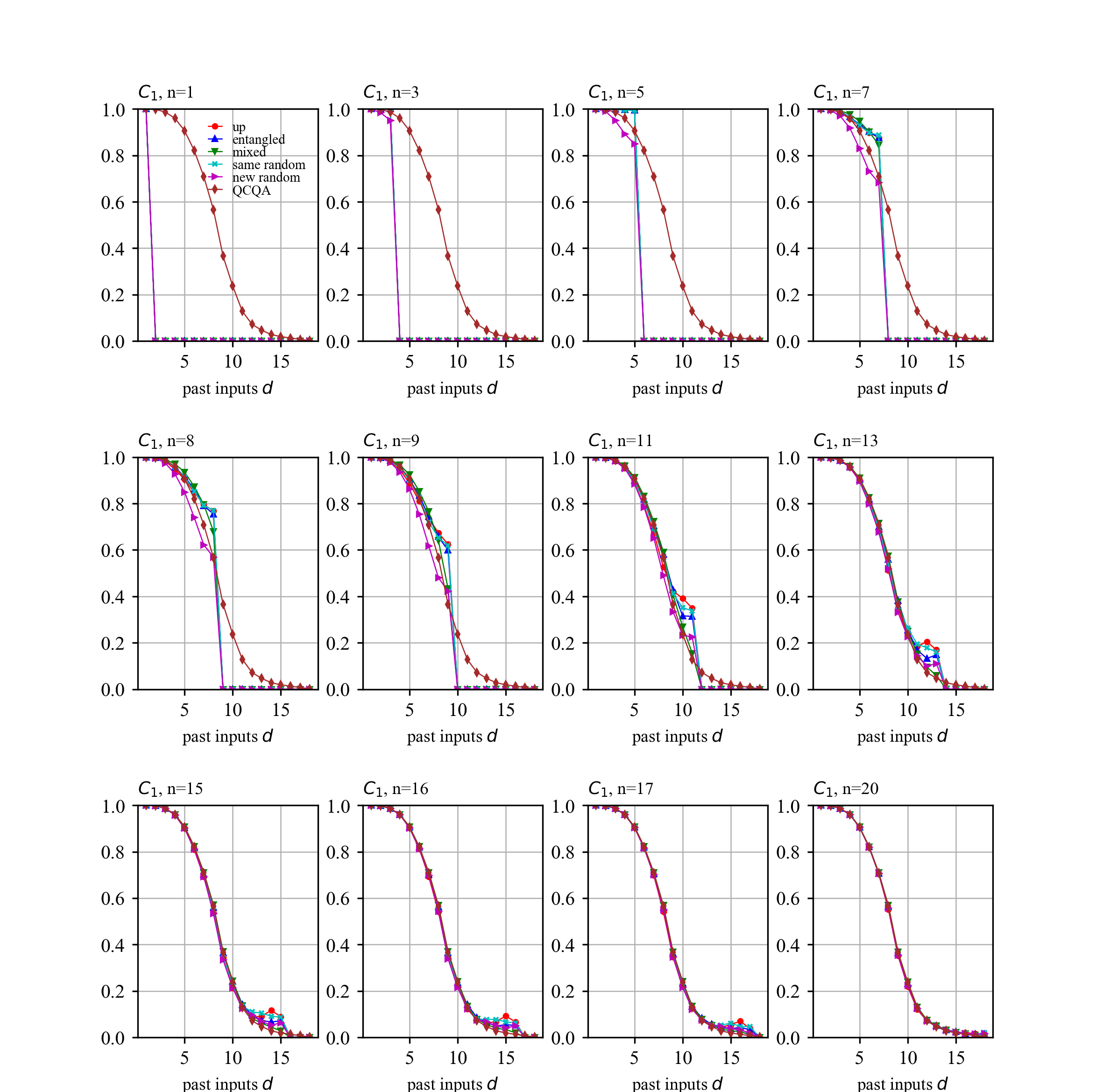}
	\caption[Image 1 ]{capacity $C_1$ for the Ising model for a different reset lengths $n$.}
\label{fig:S3}
\end{figure}

Figure \ref{fig:S2} shows the capacity for reconstructing the input $d$ steps into the past for different starting states and different schemes with a reset length of $n=15$ . The legend is explained in Table \ref{tab:states}. The up, same random, entangled, mixed and new random schemes all use the LCQA algorithm, where the starting state is changed according to the description in the table. Note that the random state are generated with "rand\_ dm" method of Qutip \cite{JoNaNo2013} with "density=1". The last scheme uses the QCQA scheme, where the previous state is reconstructed. 

\begin{table}[]
\centering
\begin{tabular}{|l|l|}
\hline
\textbf{Legend} & \textbf{Description}                                                                           \\ \hline
up              & $|0000\rangle$                                                                                       \\ \hline
same random     & The same random state as a starting state for each input.                                      \\ \hline
entangled       & ${|0000\rangle+|1111\rangle}/{\sqrt{2}}$ \\ \hline
mixed           & $\rho = Id_{16}/16$                                                                            \\ \hline
new random      & A new random state as a starting state for each input.                                         \\ \hline
QCQA            & The scheme with quadratic complexity.                                                          \\ \hline
\end{tabular}
\caption{Description of the legend. }
\label{tab:states}
\end{table}
For both the quantum circuit (Fig. \ref{fig:S2}.a) and the Ising model (Fig. \ref{fig:S2}.b), all capacitances show similar behavior up to $d=10$. 
For further past inputs, we observe that the memory capacity is higher for some of the LCQA schemes. Qualitatively, we see that states with lower von Neumann entropy (i.e. the pure states "up" and "entangled) remember their first inputs better than states with higher entropy (i.e. "same random", or with even higher entropy "new random" and "mixed").

In all schemes, the input is inserted into the first qubit. The new density matrix of all qubits is formed by the tensor product between the density matrix of the first qubit $\rho_1(i)$ and the partial trace of the density matrix of the other qubits $\text{Tr}_1(\rho(i-1))$:
\begin{align}
    \rho(i) = \rho_1(i) \otimes \text{Tr}_1{\rho(i-1)}.
\end{align}
By increasing the reset length $n$ the newly constructed density matrix $\rho(i)$ is highly likely to be a mixed state. Similarly, a random state vector is expected to be mixed. The similarities between the new random state, QCQA scheme, and the mixed state suggest that the increase around the number of the reset length $n$ of $C_1$ is influenced by two factors:
\begin{enumerate}
    \item Whether the starting state for each input is the same.
    \item Whether the starting state for each input is mixed.
\end{enumerate}
Similar behavior can be observed in the Ising model, albeit to a lesser extent. Once again, the "up", "same random", and "entangled" starting states remember early inputs better than  the "mixed" starting state, "new random" starting state, and the QCQA scheme. In this case as well, $C_1$ of the QCQA scheme exhibits almost identical behavior to the new random and mixed states.

To analyze this behavior in more detail, we will examine the behavior of $C_1$  for different values of $n$ in Figure \ref{fig:S3}. We observe that for small $n$, there is a noticeable improvement compared to the QCQA scheme, which then shifts towards the right side. For $n = 21$, no difference can be observed between the different schemes. We will attempt to quantify this behavior using additional measures. As previously discussed, the information processing capacity of the first-order IPC$_1$, often referred to as memory capacity, is given as:

\begin{align}
IPC_1 = \sum_{i_1\in comb} C_{1,1}(i_1).
\end{align}
For different $n$ we will calculate the following ratio $R$
\begin{align}
R(s) = \frac{IPC_1(i_1\leq n+1, s, LCQA)}{IPC_1(i_1\leq n+1, QCQA)},
\end{align}
the difference
\begin{align}
D(s) = {IPC_1(i_1\leq n+1, s, LCQA)}-{IPC_1(i_1\leq n+1, QCQA)}
\end{align}
and lastly a windowed difference with window length $n_W=4$
\begin{align}
D_W(s) &= {IPC_1(n+1-n_W<i_1\leq n+1, s, LCQA)} \nonumber \\
&-{IPC_1(n+1-n_W <i_1\leq n+1, QCQA)}
\end{align}

\begin{figure}[]
	\hspace*{-0.5 cm}
	\centering
	\includegraphics[scale=0.9]{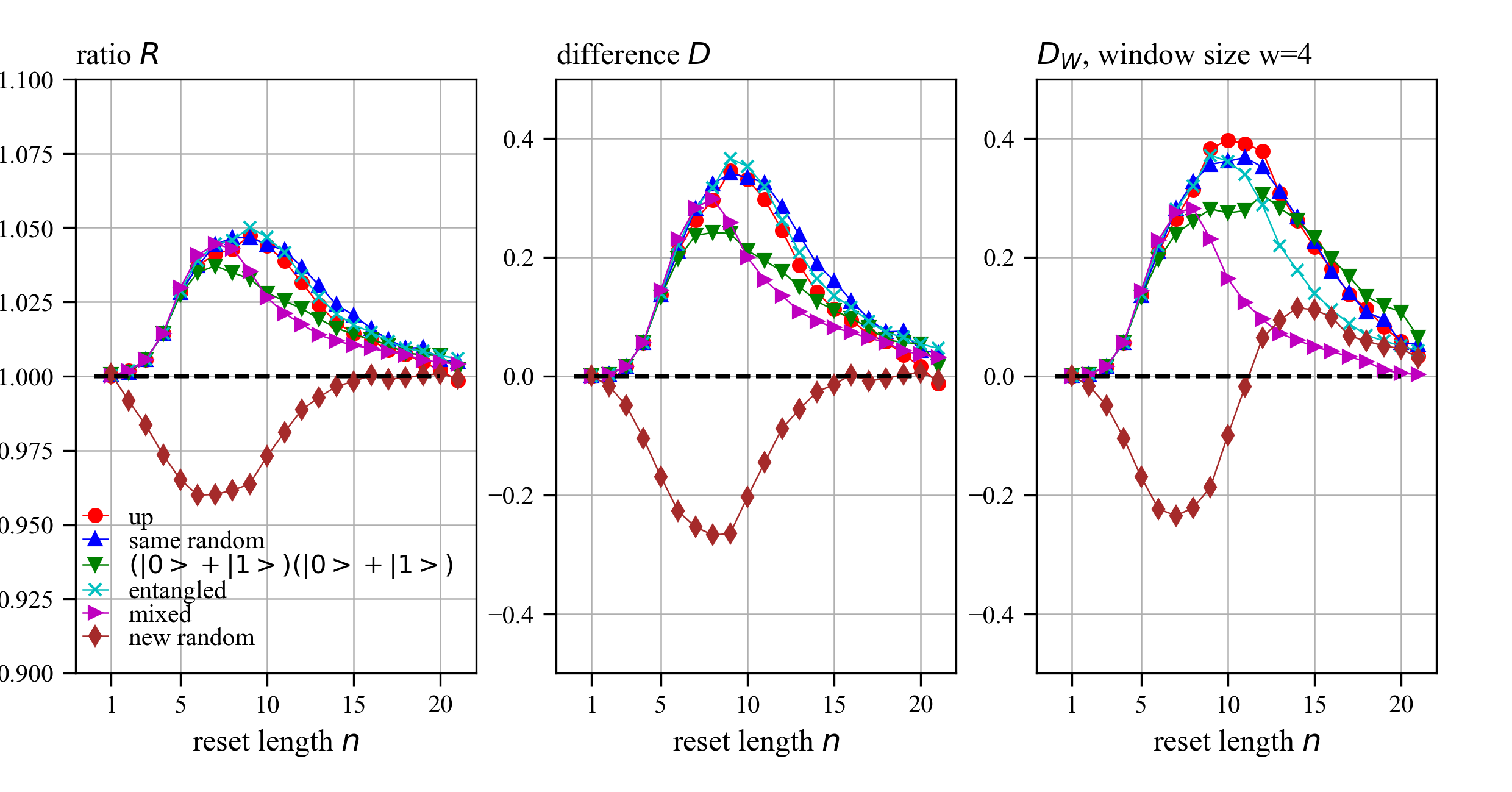}
	\caption[Image 1 ]{ratio R (left), difference D (middle) and a difference with a window size of $w=4$ over the number of reset length  $n$. }
\label{fig:S4}
\end{figure}

The horizontal line in Figure \ref{fig:S4} represents the value of the QCQA scheme. The ratio is one, while the differences become zero.

We can observe that the ratio $R$ and the difference $D$ exhibit very similar behavior. Interestingly, the mixed state performs similar to the other LCQA schemes for smaller values of $n$, but converges faster to the QCQA scheme for larger values of $n$. Furthermore, only the new random LCQA scheme shows worse performance for smaller $n$, with a ratio of $R(n=7) \approx 0.96$ and a negative difference of up to $D(n=7) \approx -0.23$.

The new random starting state initially performs worse than the QCQA scheme and saturates to the QCQA scheme for increasing $n$.

It is important to note that this analysis is not an in-depth research on the topic at hand, as the Ising model was only simulated for $T=20$, where the total information processing capacity saturates. For smaller values of $T$, a higher $IPC_1$ and smaller higher-order $IPC_{k\geq 2}$ were observed. Additionally, it would be interesting to explore additional measures beyond those presented here.
Furthermore, the behavior of $n=15$ does not have the same behaviour for the quantum circuit and Ising model. As we have observed in the previous figures, some starting states perform worse than the QCQA scheme for small values of $n$ and later they increase. 

A definitive statement on the dependence of the initial state cannot be made since only the Ising reservoir was studied in more detail. A similar behavior of $R,~D$ and $D_W$ for the quantum circuit would indicate more definite results and give rise for further research.

\bibliography{lit}